\documentclass[twocolumn,amsmath,amssymb,superscriptaddress]{revtex4}

\usepackage{graphicx}
\usepackage{dcolumn}
\usepackage{bm}
\newfont{\Fr}{eufm10}

\begin{document}


\title{Luminosity distance for Born-Infeld electromagnetic waves
propagating in a cosmological magnetic background}

\author{Mat\'\i as Aiello}
\email{aiello@iafe.uba.ar}
\thanks{ANPCyT Fellow}
\affiliation{Instituto de  Astronom\'\i a y F\'\i sica del Espacio,
Casilla de Correo 67, Sucursal 28, 1428 Buenos Aires, Argentina}
\affiliation{Departamento de F\'\i sica, Facultad de Ciencias Exactas y
Naturales, Universidad de Buenos Aires, Ciudad Universitaria, Pabell\'on
I, 1428 Buenos Aires, Argentina}
\author{Gabriel R. Bengochea}
\email{gabriel@iafe.uba.ar}
\thanks{CONICET Fellow}
\affiliation{Instituto de  Astronom\'\i a y F\'\i sica del
Espacio, Casilla de Correo 67, Sucursal 28, 1428 Buenos Aires,
Argentina} \affiliation{Departamento de F\'\i sica, Facultad de
Ciencias Exactas y Naturales, Universidad de Buenos Aires, Ciudad
Universitaria, Pabell\'on I, 1428 Buenos Aires, Argentina}
\author{Rafael Ferraro}
\email{ferraro@iafe.uba.ar}
\thanks{Member of Carrera del Investigador Cient\'{\i}fico (CONICET,
Argentina)}
\affiliation{Instituto de  Astronom\'\i a y F\'\i sica del Espacio,
Casilla de Correo 67, Sucursal 28, 1428 Buenos Aires, Argentina}
\affiliation{Departamento de F\'\i sica, Facultad de Ciencias Exactas y
Naturales, Universidad de Buenos Aires, Ciudad Universitaria, Pabell\'on
I, 1428 Buenos Aires, Argentina}


\begin{abstract}
\begin{center}
\end{center}

Born-Infeld electromagnetic waves interacting with a static
magnetic background are studied in an expanding universe. The
non-linear character of Born-Infeld electrodynamics modifies the
relation between the energy flux and the distance to the source,
which gains a new dependence on the redshift that is governed by
the background field. We compute the luminosity distance as a
function of the redshift and compare with Maxwellian curves for
supernovae type Ia.

\end{abstract}

\pacs{Valid PACS appear here}
\maketitle

\section{Introduction}

The discovery of an unexpected diminution in the observed energy
fluxes coming from supernovae type Ia \cite{Riess-Perlmutter},
which are thought of as standard candles, has been interpreted in
the context of the standard cosmological model as evidence for an
accelerating universe dominated by something called dark energy.
This fact is one of the most puzzling and deepest problems in
cosmology and fundamental physics today. Although the cosmological
constant seems to be the simplest explanation for the phenomenon,
several dynamical scenarios have been tried out (see, for
instance, \cite{padmanabhan} and references therein). It is
worthwhile to emphasize that the evidence for an accelerating
universe mainly relies on energy flux measurements for type Ia
supernovae at different values of cosmological redshifts. They
provide the most direct and consistent way to determine the recent
expansion history of the universe. Nevertheless the relation
between the cosmological redshift and the energy flux for a
point-like source involves not only the evolution of the universe
during the light journey but some assumptions about the nature of
the light itself. Customarily, one accepts the linear Maxwell
theory to describe the light propagation, where light propagates
without interacting with other electric or magnetic fields.
However, in the context of non-linear electrodynamics the
interaction between the light emitted from such distant sources
and cosmological magnetic backgrounds modifies the relation
between the redshift and the flux of energy. If this kind of
effect were not correctly interpreted it could lead to an
erroneous conclusion about the expansion history of the universe.
Concretely, an effect coming from non-linear electrodynamics could
explain the curves of luminosity distance vs redshift for type Ia
supernovae without invoking dark energy. This remark drives us to
study the propagation of non-linear electromagnetic waves in an
expanding universe with a magnetic background. We will benefit
from recently obtained results for Born-Infeld electromagnetic
plane waves propagating in a magnetic uniform background in
Minkowski space-time \cite{9abf}.

Born-Infeld electrodynamics \cite{1born,2born} is a non-linear
theory for the electromagnetic field $F_{\mu\nu}$ governed by the
Lagrangian
\begin{equation}
{\cal L}=\sqrt{-g}\
\frac{b^2}{4\:\pi}\,\left(1-\sqrt{1+\frac{2S}{b^2}-
\frac{P^2}{b^4}}\right)
\end{equation}
where $b$ is a new fundamental constant and $S$ and $P$ are the
scalar and pseudoscalar field invariants
\begin{equation}
S=\frac{1}{4}F_{\mu\nu}F^{\mu\nu}\ \ \ \ \ \ \ \ \ \
P=\frac{1}{4}\:^{*}F_{\mu\nu}F^{\mu\nu}
\end{equation}
(in Minkowski space-time it is $2S=|{\bf{\mathbb
B}}|^2-|{\bf{\mathbb E}}|^2$ and $P= {\bf{\mathbb
E}}\cdot{\bf{\mathbb B}}$, $\bf{\mathbb E}$ and $\bf{\mathbb B}$
being the electric and magnetic fields respectively). Born-Infeld
electrodynamics goes to Maxwell electromagnetism when
$b\rightarrow \infty$. In particular, the Born-Infeld field of a
point-like charge reaches the finite value $b$ at the charge
position and becomes Coulombian far from the charge. Born-Infeld
theory is the only non-linear spin 1 field theory displaying
causal propagation and absence of birefringence \cite{deser,5boi}.

Nowadays, Born-Infeld theory is reborn in the context of
superstrings because Born-Infeld-like Lagrangians emerge in the
low energy limit of string theories \cite{strings}.
Born-Infeld-like Lagrangians have been also proposed to describe a
matter dynamics able to drive the universe to an accelerated
expansion \cite{varios}. In spite of this revival of Born-Infeld's
ideas, there is no experimental evidence for Born-Infeld effects
in electrodynamics: the value of $b$ remains unknown (see an upper
bound for $b$ in \cite{Jack}).

In the next section we will summarize the recently obtained
results on Born-Infeld waves propagating in Minkowski space-time
in the presence of a uniform magnetic background \cite{9abf}.
These results --properly adapted-- will be used in section III to
understand the Born-Infeld energy flux coming from a point-like
source in a spatially flat Friedman-Robertson-Walker (FRW)
expanding universe. In section IV we will reformulate the relation
between the luminosity distance $d_L$ and the redshift $z$ within
the framework of Born-Infeld electrodynamics. We will show that
the presence of magnetic backgrounds modify the curves $d_L$ vs
$z$, and we will analyze the consequences for the measurements of
the luminosity distance of supernovae type Ia.

\section{Born-Infeld Plane waves in Minkowski space-time revisited}
Free Born-Infeld electromagnetic plane waves do not differ from
Maxwell plane waves. However, if the wave propagates in the
presence of background fields then the propagation velocity
becomes lower than $c$ as a consequence of the non-linearity of
the theory. This issue has been studied in
\cite{5boi,3pleb,4pleb,6novello,7novello,8novello} by considering
the propagation of discontinuities. Recently the exact solution
for a Born-Infeld electromagnetic plane wave propagating in a
static uniform magnetic background has been obtained \cite{9abf}.
The result retrieves the value for the phase velocity obtained in
the above mentioned references:
\begin{equation}
\beta\ =\
\left(\frac{1+\frac{B_L^2}{b^2}}{1+\frac{B^2}{b^2}}\right)^{\frac{1}{2}}
\label{beta}
\end{equation}
where ${\bf B}=B_L\, \hat{x} + B_B\, \hat{y} + B_E\, \hat{z}$ is
the background field and $B_L$ is its component along the
propagation direction.

Furthermore, the results of \cite{9abf} exhibit an up to now
unknown feature: the wave develops a longitudinal electric field
which depends on $B_L$ and $B_E$ (the projection of the background
field ${\bf B}$ on the polarization direction). In fact, the total
electric and magnetic fields result
\begin{eqnarray}\nonumber &&
{\bf{\mathbb E}}= \beta\ E_{wave}(\xi)\,\hat{z}+{\bf
E}_{long}(\xi) \ ,\\ && {\bf{\mathbb B}}= -E_{wave}(\xi)\,\hat{y}
+ {\bf B}\label{field}
\end{eqnarray}
where $E_{wave}(\xi)$ is the (usual) transversal electric field of
the wave, and $\xi=x-\beta\, t$ is its phase. The longitudinal
electric field of the wave is
\begin{eqnarray} \nonumber
{\bf E}_{long} &&= \frac{-{\bf B}\cdot \big( E_{wave}(\xi)\,
\hat{z} \big)\ }{b^2\,\Big(1+\frac{B_L^2}{b^2}\Big)^{\frac{1}{2}}
\Big(1+\frac{B^2}{b^2}\Big)^{\frac{1}{2}}}\ {\bf B}_L \\ && =
\frac{-B_E B_L \, E_{wave}(\xi)\
}{b^2\,\Big(1+\frac{B_L^2}{b^2}\Big)^{\frac{1}{2}}
\Big(1+\frac{B^2}{b^2}\Big)^{\frac{1}{2}}}\ \hat{x}\ \label{elong}
\end{eqnarray}
(of course ${\bf E}_{long} $ vanishes when $b\rightarrow\infty$).
As a consequence, the Poynting vector ${\bf\cal S}$ acquires a
component transverse to the propagation direction. In Born-Infeld
electrodynamics the Poynting vector results
\begin{equation}
{\bf\cal S}=\frac{1}{4\, \pi}\, \frac{{\bf{\mathbb E}}\times
{\bf{\mathbb B}}}{\sqrt{1+ b^{-2}\, (|{\bf{\mathbb
B}}|^2-|{\bf{\mathbb E}}|^2)- b^{-4}\, ({\bf{\mathbb
E}}\cdot{\bf{\mathbb B}})^2}}
\end{equation}

For a wave $E_{wave}=E_o \cos(x-\beta t)$ the time averaged values
of the transversal and longitudinal components of the Poynting
vector associated with the field (\ref{field}) are
\begin{equation}
<{\bf\cal S}_\perp>\ =\ \frac{E_o^2\, B_L\, {\bf B_\perp}}{8\,
\pi\, b^2}\ +\ {\cal O}(b^{-4})\label{poynting1}
\end{equation}
\begin{equation}
<{\cal S}_x>\ =\ \frac{E_o^2}{8\, \pi}\left[1\ -\ \frac{4
B_\perp^2-2 B_E^2+B_L^2}{2\, b^2}\right]\ +\ {\cal
O}(b^{-4})\label{poynting2}
\end{equation}
Notice that the transversal part of $<\pmb{\cal S}>$ is parallel
to the transversal background field ${\bf B_\perp}$. The angle
$\alpha$ between the ray and the propagation direction is
\begin{equation}
\tan\alpha\ =\ \frac{B_L\ \vert {\bf B}_\perp\vert}{b^2}\ +\ {\cal
O}(b^{-4})\label{alpha}
\end{equation}
Although this effect resembles the behavior of the extraordinary
ray in anisotropic media, it should be emphasized that no
birefringence exists in this case, which is a distinctive feature
of Born-Infeld non-linear electrodynamics \cite{5boi,6novello}.

In Born-Infeld electrodynamics the energy density is
\begin{equation}
\rho = \frac{b^2}{4\ \pi}\left[\frac{1\ +\ b^{-2}\, |{\bf{\mathbb
B}}|^2}{\sqrt{1+ b^{-2}\, (|{\bf{\mathbb B}}|^2-|{\bf{\mathbb
E}}|^2)- b^{-4}\, ({\bf{\mathbb E}}\cdot{\bf{\mathbb
B}})^2}}-1\right]
\end{equation}
Then, the time averaged energy density associated with the wave is
\begin{eqnarray}\nonumber
<\rho>\ \ =&& \frac{E_o^2}{8 \pi}\bigg[ 1-\frac{3
B_\bot^2-2B_E^2+B_L^2}{2\, b^2} \bigg]\\ && + \frac{{B}^2}{8\,
\pi} \bigg[1 -\frac{{B}^2}{4 b^2} \bigg]+\, {\cal O}(b^{-4})
\label{rho}
\end{eqnarray}
The above enumerated properties provides the rules to propagate
Born-Infeld light rays in the presence of static uniform magnetic
backgrounds.

\section{Born-Infeld Plane waves in an expanding flat FRW universe}
In this section we will study the energy flux of Born-Infeld waves
propagating in a spatially flat FRW expanding universe,
\begin{equation}\label{metric}
ds^2=a(\eta)^2(d\eta^2-dx^2-dy^2-dz^2)
\end{equation}
The conformal time $\eta$ is related with the cosmological time
$t$ (the proper time of the comoving fluid) through the equation
$dt=a(\eta)\, d\eta$. The energy-momentum conservation laws,
$T^{\mu\nu}_{\ \ \ ;\nu}=0$, can be written as
\begin{equation}
\big[\sqrt{-g}\:T^{\mu\nu}\big]_{,\nu}+\Gamma^{\mu}_{\nu\rho}\:\sqrt{-g}\:T^
{\rho\nu}=0
\end{equation}
For the geometry (\ref{metric}) it is $\Gamma^0_{\nu\rho}=a_{,0}\,
a^{-1}\, \delta_{\nu\rho}$, thus the energy balance becomes
\begin{equation}
a^{-2}\, \big[a^4\,T_0^0\big]_{,0} + a^2\,
{T_0^\alpha}_{,\alpha}-a_{,0}\, a\ T = 0\label{conservacion}
\end{equation}
where $\alpha=1,2,3$ and $T=T_\mu^{\; \mu}$ is the trace of the
energy-momentum tensor. We will use (\ref{conservacion}) to study
electromagnetic waves propagating in a magnetic background; so,
$T^\nu_0$ in (\ref{conservacion}) includes both the wave and the
background fields. In Maxwell electromagnetism the trace $T$ is
identically null. As a consequence, if $\{T_0^\nu\}^M$ solves
(\ref{conservacion}) in Minkowski space-time ($a=constant$) then
$\{T_0^\nu\}^{FRW} = a(\eta)^{-4}\, \{T_0^\nu\}^M$ will solve
(\ref{conservacion}) in the spatially flat FRW universe. Since
Maxwell energy-momentum tensor is quadratic in the fields, the
former assertion means that the Maxwell fields scale with the
factor $a(\eta)^{-2}$. This general conclusion is applicable to
the particular case of a wave propagating in a magnetic
background.

On the contrary, $T$ is non-null in Born-Infeld electrodynamics.
Thus, the scaling of $T_0^\nu$ with the factor $a(\eta)^{-4}$ does
not guarantee the fulfillment of (\ref{conservacion}). An
additional correction is needed, which must vanish in the limit
$b\rightarrow\infty$. We will search this correction to the lowest
order in $b^{-2}$ for the Born-Infeld plane wave propagating in a
magnetic background, whose main features were depicted in section
II. Of course, we will assume that the magnetic background do not
sensitively affects the homogeneity and isotropy of the space-time
dominated by matter and (presumably) dark energy, because the
energy density of the background field is negligible compared with
matter and cosmological constant densities (a typical value for
this field is $10^{-7} G$).

Since we are going to consider the lowest order correction in
$b^{-2}$ for the magnitudes described in section II such as
(\ref{elong}), (\ref{poynting1}), (\ref{poynting2}) and
(\ref{rho}), we remark that any magnitude of this order or bigger
will only require the Maxwellian scaling of the fields with the
factor $a(\eta)^{-2}$. In the case of the propagation velocity
(\ref{beta}) the correction can also be obtained from the results
in \cite{3pleb,6novello}, where it is shown that the equation
accomplished by the wave four-vector can be understood as if the
rays propagate along null geodesics of an effective metric
$\bar{g}_{\mu\nu}$. In the case of the Born-Infeld electrodynamics
the effective metric $\bar{g}_{\mu\nu}$ is:
\begin{equation} \bar{g}^{\mu\nu}=(b^2+\frac{1}{2}F_{\rho \sigma} F^{\rho
\sigma}) g^{\mu\nu}+ F^{\mu}_{\ \lambda} F^{\lambda \nu}
\label{geomef}
\end{equation}
where $g_{\mu\nu}$ is the space-time metric and $F_{\mu \nu}$ is
the electromagnetic background where the ray propagates, in this
case the components of the background are,

\begin{equation}
F_{x y}=-B_E,\:\: F_{y z}=-B_L,\:\: F_{x z}=B_B \label{campob}
\end{equation}

For a ray propagating along the $x$ direction it is
\begin{equation} d\bar{s}^2=\bar{g}_{\eta\eta}d\eta^2+\bar{g}_{xx}dx^2=0 \Rightarrow
\frac{dx}{d\eta}=\sqrt{-\frac{\bar{g}_{\eta\eta}}{\bar{g}_{xx}}}
\label{geoefe} \end{equation} When the effective metric
(\ref{geomef}) is evaluated for the magnetic background
(\ref{campob}) it results
\begin{equation} \bar{g}_{\eta\eta}=\frac{1}{\bigg(b^2+\frac{\textbf{B}^2}{a(\eta)^4}\bigg) a(\eta)^2} \:\:\:\:\:\:\:\:\:\:\:\:\:
\bar{g}_{xx}=\frac{-1}{\bigg(b^2+\frac{B^2_L}{a(\eta)^4}\bigg)a(\eta)^2}
\end{equation}
Thus, it is obtained
\begin{equation}
\beta(\eta)\ =\frac{dx}{d\eta}=\
\left(\frac{1+\frac{B_L^2}{a(\eta)^4 b^2 }}{1+\frac{B^2}{a(\eta)^4
b^2}}\right)^{\frac{1}{2}} \label{betame}
\end{equation}
By integrating the ray path in (\ref{geoefe}) we recognize that
the phase $\xi$ has to be replaced with
\begin{equation}
\xi\ =\ x\ -\ \int^\eta \beta(\eta')\ d\eta'\label{phase}
\end{equation}
(compare with the adiabatic treatment for an oscillator with
slowly variable frequency \cite{goldstein,birrel}). Notice that,
since $\partial\xi/\partial x=1$ and
$\partial\xi/\partial\eta=-\beta$ as in Minkowski space-time, then
the derivatives of $T^\mu_\nu$ in (\ref{conservacion}) will
preserve their Minkowskian structure. The trace of the Born-Infeld
energy-momentum tensor is (see for instance \cite{9abf})
\begin{equation}
T=\frac{b^2}{\pi}\left[\frac{1+b^{-2}\,S}{\sqrt{1+\frac{2
S}{b^2}-\frac{P^2}{b^4}}}-1\right]=\frac{S^2+P^2}{2\pi\,
b^2}+{\cal O}(b^{-4}) \label{trace}
\end{equation}
and scales with $a(\eta)^{-8}$. Let us consider a solution
$\{T_0^\nu\}^M$ for the Born-Infeld field in Minkowski space-time;
i.e., $\{T_0^0\}^M_{\ \ ,0} + \{T_0^\alpha\}^M_{\ \ ,\alpha} = 0$
(for the case under consideration, $\{T_0^0\}^M$ is the energy
density (\ref{rho}), and $\{T_0^\alpha\}^M$ are the components of
the Poynting vector (\ref{poynting1}-\ref{poynting2})). As stated
above, the scaling of the fields with $a(\eta)^{-2}$ is not enough
to get a solution for (\ref{conservacion}), since the term
associated with the trace is now non-null. As the trace is of
order $b^{-2}$, only $T_0^\nu$ has to be corrected in
(\ref{conservacion}). So, let us try with the scaling
$\{T_0^\nu\}^{FRW}=(1-\varepsilon\, a^{-4}\,b^{-2})\,a^{-4}\,
\{T_0^\nu\}^M$. By replacing in (\ref{conservacion}) we obtain
\begin{eqnarray}
\nonumber a^{-2}\,(1-\varepsilon\,a^{-4}\,b^{-2})(\{T_0^0\}^M_{\ \
,0} + \{T_0^\alpha\}^M_{\ \ ,\alpha})\\
+4\,\varepsilon\,b^{-2}\,a_{,0}\,a^{-7}\, \{T_0^0\}^M-a_{,0}\,
a^{-7}\ \{T\}^M = 0
\end{eqnarray}
Therefore
\begin{equation}
\varepsilon=b^2 \frac{\{T\}^M}{4\, \{T_0^0\}^M}\label{epsilon}
\end{equation}
Since the trace is of order $b^{-2}$, then $S$, $P$ and
$\rho=T_0^0$ can be computed with the Maxwellian fields. The time
averaged trace for a plane wave traveling in a magnetic background
is \cite{9abf}
\begin{equation}
<\{T\}^M>\ =\ \frac{B^4+2 E_o^2B_\bot^2}{8\pi
b^2}\label{averagedtrace}
\end{equation}
Thus $\varepsilon$ results
\begin{equation}
\varepsilon=\frac{B^4+2 E_o^2B_\bot^2}{4
(B^2+E_o^2)}\label{epsilon}
\end{equation}

\section{Luminosity distance for a point-like source}
According to the results of previous section, we should correct
the Minkowskian Poynting vector (\ref{poynting1}),
(\ref{poynting2}) with the factor $(1-\varepsilon\,
a^{-4}\,b^{-2})\,a^{-4}$ \cite{pie1}. Thus the energy flux along
the propagation direction is
\begin{eqnarray}
\nonumber <{\cal S}_x>\ =\ \left(1-\frac{\varepsilon}{a(\eta)^4\,
b^2}\right)\frac{E_o^2}{8\, \pi\, a(\eta)^4}\\ \nonumber\left[1\
-\ \frac{4 B_\perp^2-2 B_E^2+B_L^2}{2\,a(\eta)^4 b^2}\right]\ +\
{\cal O}(b^{-4})\\ =\ \left(1-\frac{\Delta_B}{a(\eta)^4\,
b^2}\right)\frac{E_o^2}{8\, \pi\, a(\eta)^4}\ +\ {\cal O}(b^{-4})
\label{poyntingFRW}
\end{eqnarray}
where $\Delta_B= 2 B_{\bot}^2-B_E^2+\frac{B_L^2}{2}+\frac{B^4+2
E_0^2 B_{\bot}^2}{4(B^2+E_0^2)}$ is positive. Actually we are
interested in light rays emitted from point-like sources in the
universe. When a light ray travels in the intergalactic space, it
undergoes a background magnetic field in different regions of the
traveled path. In such case, the values of $B$, $B_E$, etc.
implied in the propagation should be taken as representative
values of the intergalactic background field along the ray
trajectory; so $\Delta_B$ is a magnitude of the same order as the
representative squared intergalactic field.

For a radial propagation the Minkowskian spherical energy flux is
obtained from the plane flux by dividing by the square of the
radial distance to the source $r^2=x^2+y^2+z^2$. Therefore the
radial energy flux in a FRW spatially flat universe becomes
\begin{equation}
\mbox{\Fr F}\ =\ \left(1-\frac{\Delta_B}{a(\eta)^4\,b^2}\right)
\frac{E_o^2}{8\, \pi\,a(\eta)^4\,r^{2}}\ +\ {\cal
O}(b^{-4})\label{tetarpro}
\end{equation}
In Maxwell electrodynamics the background field does not interfere
with the wave, but in non-linear Born-Infeld electrodynamics the
first correction in (\ref{tetarpro}) displays a coupling between
the wave and the background magnetic field.
 The luminosity
of an object results from integrating the flux at the time of
emision $\eta_e$, this integration normalizes $E_o$ to fit the
value of the luminosity $L$. By using (\ref{tetarpro}) one obtains
\begin{equation}\label{lum2}
L =\ \frac{E_o^2}{2\:a_e^2}\, \left(1-\frac{\Delta_B}{a_e^4\,
b^2}\right)
\end{equation}
Combining (\ref{tetarpro}) and (\ref{lum2}) the energy flux can be
written as
\begin{equation}\label{fluxpro2}
\mbox{\Fr F}\ \simeq\ \frac{L\, a_e^2}{4\pi\:a(\eta)^4
r^2}\left[1+ \frac{\Delta_B}{a(\eta)^4\,
b^2}\left(\frac{a(\eta)^4}{a_e^4}-1\right)\right]
\end{equation}
The luminosity distance $d_L$ is defined as (see for instance
\cite{Turner})
\begin{equation}\label{defdl}
{d_L}^2\ \equiv\ \frac{L}{4\:\pi\: {\mbox{\Fr F}}_{ o}}
\end{equation}
where $\mbox{\Fr F}_{ o}$ is the flux measured at time $\eta_o$ at the
position of the observer $r_o$ (the source is at $r=0$). Therefore
\begin{equation}\label{dl}
d_L\, =\, \frac{a_o^2\:r_o}{a_e} \left[1- \frac{\Delta_B}{2\,
a_o^4 b^2}\left(\frac{a_o^4}{a_e^4}-1\right)\right]\ +\ {\cal
O}(b^{-4})
\end{equation}
In order to relate the luminosity distance with the redshift let
us consider the motion of a ray: $dr = \beta(\eta)\, d\eta$. So,
if a ray is emitted at time $\eta_e$ from a source located at
$r=0$, and arrives at time $\eta_o$ to the position $r_o$ of the
observer, then the wavecrest emitted at time $\eta_e+\delta\eta_e$
will arrive at the observer at time $\eta_o+\delta\eta_o$, in such
a way that
\begin{equation}\label{int1}
\int_{\eta_e}^{\eta_o}
\beta\:d\eta=\int_{\eta_e+\delta\eta_e}^{\eta_o+\delta\eta_o}
\beta\:d\eta
\end{equation}
or, equivalently
\begin{equation}\label{int2}
\int_{\eta_e}^{\eta_e+\delta\eta_e}\beta\:d\eta=\int_{\eta_o}^{\eta_o+\delta\eta_o}
\beta\:d\eta
\end{equation}
Thus $\beta_e\, \delta\eta_e = \beta_o\, \delta\eta_o$, i.e.
$\beta_e\, a^{-1}_e\, \delta t_e = \beta_o\, a_o^{-1}\,
\delta t_o$. therefore the redshift is
\begin{eqnarray}\label{redshift}
\nonumber &&1 + z\equiv \frac{\nu_e}{\nu_o}=\frac{\delta
t_o}{\delta t_e}\\ &&=\,
\frac{a_o}{a_e}\:\bigg[1+\frac{B_{\bot}^2}{2\:a_o^4
b^2}\left(1-\frac{a_o^4}{a_e^4}\right)\bigg]\ +\ {\cal O}(b^{-4})
\end{eqnarray}
By inverting this relation one obtains
\begin{equation}\label{ao/ae}
\frac{a_o}{a_e}\simeq\, \big(1+z\big)\:
\bigg[1+\Big[\big(1+z\big)^4-1\Big]\:\frac{B_{\bot}^2}{2 a_o^4
b^2}\bigg]
\end{equation}
This quotient is one of the components in the luminosity distance
(\ref{dl}). The other one is the proper distance $a_o r_o$. This
distance depends on how the universe evolves. In fact, following
the motion of the ray, it results
\begin{equation}
a_o\, r_o = \int_{\eta_e}^{\eta_o} a_o\, \beta(\eta)\, d\eta =\int_z^0
\frac{a_o\, \beta(z')}{a(z')\, \frac{dz'}{dt}}\, dz'
\end{equation}
where $z'$ is the redshift of a wave emitted at time $t \leq t_o$.
One can replace $dz'/dt$ in terms of the Hubble parameter
$H\equiv\frac{\dot
a}{a}=\frac{d}{dt}\:\log\big(\frac{a(t)}{a_o}\big)$ by performing
the derivative of the logarithm of (\ref{ao/ae}),
\begin{equation}\label{H}
H(z)\simeq\, -\frac{1}{(1+z)}\:\bigg[1+\big(1+z\big)^4\: \frac{2
B_{\bot}^2}{a_o^4 b^2}\bigg]\:\frac{dz}{dt}
\end{equation}
Thus
\begin{eqnarray}
\nonumber &&a_o\, r_o \simeq \int_0^z
\frac{a_o}{a(z')}\:\bigg[1+\big(1+z'\big)^4\: \frac{2
B_{\bot}^2}{a_o^4 b^2}\bigg]\: \frac{\beta(z')\,
dz'}{(1+z')H(z')}\\ \nonumber &&\simeq \int_0^z
\frac{a_o}{a(z')}\:\bigg[1+\big(1+z'\big)^4\: \frac{3 B_{\bot}^2}{2 a_o^4 b^2}\bigg]\: \frac{dz'}{(1+z')H(z')}\\
&&\simeq \int_0^z \bigg[1+\big(4(1+z')^4-1\big)\:
\frac{B_{\bot}^2}{2 a_o^4 b^2}\bigg]\:
\frac{dz'}{H(z')}\label{aoro}
\end{eqnarray}
Einstein equations say that the Hubble parameter for a spatially
flat universe dominated by matter and cosmological constant is
\cite{Turner}
\begin{equation}\label{H2}
H(z)^2=H_o^2\:\bigg[\Omega_m\:\left(\frac{a_o}{a}\right)^3+\Omega_\Lambda\bigg]
\end{equation}
where $\Omega_m$ and $\Omega_\Lambda$ are the contributions from
matter and cosmological constant to the total density of the
universe (so we are considering here $\Omega_m+\Omega_\Lambda=
1$). From (\ref{ao/ae}) one knows that
\begin{equation}\label{H3}
\frac{H(z)^2}{H_o^2}\simeq
\Omega_m(1+z)^3\bigg[1+\big[(1+z)^4-1\big]\frac{3 B_{\bot}^2}{2
a_o^4 b^2}\bigg]+\Omega_\Lambda
\end{equation}
Therefore the proper distance (\ref{aoro}) times the present
Hubble parameter results
\begin{eqnarray}
\nonumber &&H_o\, a_o\, r_o\simeq\int_1^{1+z}\, dZ\,
\Bigg[\frac{1}{\sqrt{\Omega_\Lambda+\Omega_mZ^3}}+\\
\nonumber&&+\frac{3 B_{\bot}^2}{2 a_o^4
b^2}\left(\frac{4Z^4-1}{3\sqrt{\Omega_\Lambda+\Omega_mZ^3}}
-\frac{\Omega_mZ^3\left(Z^4-1\right)}{2\left(\Omega_\Lambda+
\Omega_mZ^3\right)^{\frac{3}{2}}}\right)\Bigg]\\
\end{eqnarray}
By replacing this integral in (\ref{dl}) and combining it with
(\ref{ao/ae}), the luminosity distance turns out to be

\begin{eqnarray}\label{Hodl}
\nonumber&& H_o\, d_L \simeq (1+z) f_L(z)+ \frac{3 B_{\bot}^2}{2
a_o^4 b^2} (1+z) g_L(z)+ \\&& +\Bigg(
\frac{B_{\bot}^2-\Delta_{B}}{2 a_o^4 b^2} \Bigg) \,
(1+z)\left[(1+z)^4-1\right] f_L(z)
\end{eqnarray}
where
\begin{equation} \label{Hod3}
g_{L}(z)=\int_1^{1+z}\left[\frac{4Z^4-1}{3\sqrt{\Omega_\Lambda+\Omega_mZ^3}}-
\frac{\Omega_mZ^3\left(Z^4-1\right)}{2\left(\Omega_\Lambda+
\Omega_mZ^3\right)^{\frac{3}{2}}}\right]dZ
\end{equation}
and
\begin{equation}\label{Hodl2}
f_L(z)= \int_1^{1+z}\frac{1}{\sqrt{\Omega_\Lambda +\Omega_mZ^3}}\,
dZ \ \ \ \ \ \
\end{equation}

In (\ref{Hodl}) the last two terms containing  $f_L$ and $g_L$
characterizes the correction to the luminosity distance coming
from the background magnetic field at the considered order
\cite{pie2}. The factor multiplying the third term,
$\frac{B_{\bot}^2-\Delta_{B}}{2 a_o^4 b^2}$, is negative. The
functions $f_L(z)$ and $g_L(z)$ depends on $\Omega_m$ and
$\Omega_\Lambda$. In order to analyze the extra terms in the
luminosity distance we will display the case where the three
components of the representative magnetic background are equal.
Calling these components $B$ and replacing in (\ref{Hodl}) we
obtain:

\begin{equation}\label{Hod5}
H_o\, d_L \simeq (1+z) f_L(z)+ \frac{B^2}{2 a_o^4 b^2} F(z)
\end{equation}
where
\begin{equation}
F(z)= 6 \, (1+z) g_L(z)- \frac{9}{2}(1+z)\left[(1+z)^4-1\right]
f_L(z)
\end{equation}

\begin{figure}
\begin{center}
\includegraphics[width=8cm,angle=0]{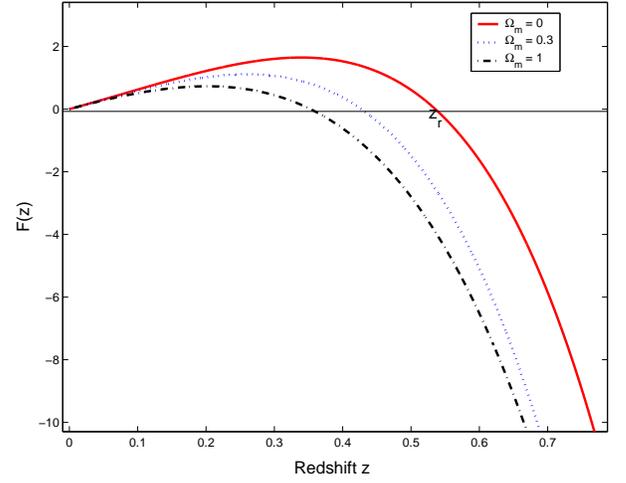}
\caption[]{Luminosity distance correction $F(z)$  for three
different sets of $(\Omega_m, \Omega_\Lambda)$. If $z>z_r$ the
correction becomes negative ($z_r$ is the zero of $F(z)$).}
\label{correccion}
\end{center}
\end{figure}

Figure \ref{correccion} shows the  behavior of $F(z)$ for three
different models. If $\Omega_m=0$ then $F(z)$ has a maximum at
$z=0.35$  and becomes negative when $z
> z_r = 0.54$, if $\Omega_m=0.3$ $F(z)$ has a maximum at $z=0.26$ and becomes negative when $z
> z_r = 0.43$ and if $\Omega_m=1$ then $F(z)$ has a maximum
at $z=0.21$ and becomes negative when $z
> z_r = 0.35$.  It
can be seen that the lower $\Omega_m$ the higher $z_r$.

\begin{figure}
\begin{center}
\includegraphics[width=8cm,angle=0]{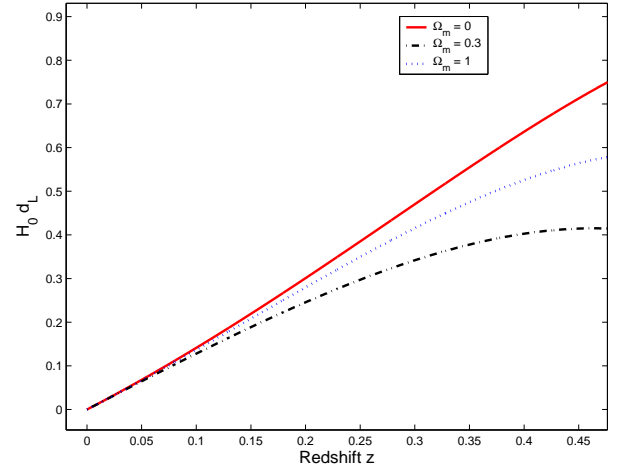}
\caption[]{Luminosity distance times the present Hubble parameter
for three different sets of $(\Omega_m, \Omega_\Lambda)$ ($ \frac{
B^2}{ 2 a_o^{4}b^{2}}=5\times10^{-3}$).} \label{distancia}
\end{center}
\end{figure}

\begin{figure}
\begin{center}
\includegraphics[width=8cm,angle=0]{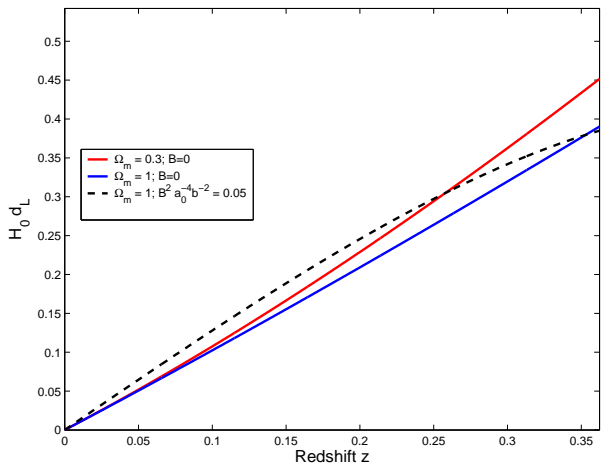}
\caption[]{Low redshift behaviors of ordinary (Maxwellian) cosmologies
(solid lines) compared with Born-Infeld electrodynamics models without
cosmological constant.} \label{onlymatter}
\end{center}
\end{figure}

\begin{figure}
\begin{center}
\includegraphics[width=8cm,angle=0]{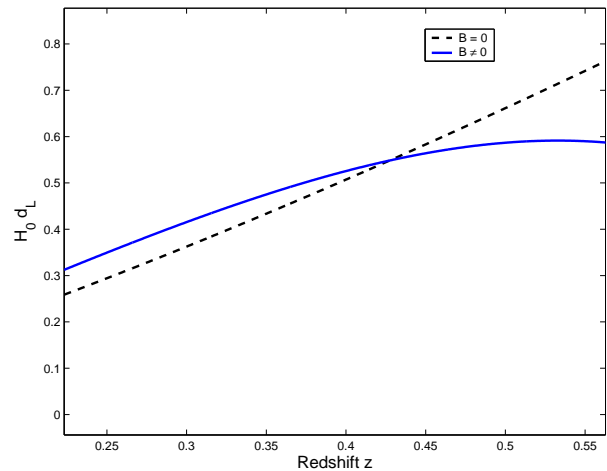}
\caption[]{Luminosity distance times the present Hubble parameter
for $\Omega_m=0.3$, $\Omega_\Lambda=0.7$. Comparison between
Maxwell (dashed line) and Born-Infeld (solid line) electrodynamics
($\frac{B^2}{2 a_o^{4}b^{2}}=5\times10^{-3}$).}
\label{comparacion}
\end{center}
\end{figure}

 Figure \ref{distancia} displays the luminosity distance times
the present Hubble parameter (\ref{Hodl}) for the three cases
considered in Fig.\ref{correccion} (we have chosen $\frac{B^2}{ 2
a_o^{4}b^{2}}=5\times 10^{-3}$). When $z\ll1$ the leading terms in
(\ref{Hod5}) are $H_o d_L\simeq z\,(1+ \frac{3
B_{\bot}^2}{a_o^{4}b^{2}})$. So the slope at low redshift gets a
contribution coming from the background field.
Figure \ref{onlymatter} compares the low redshift behavior of
standard cosmology ($\Omega_\Lambda = 0.7$, $\Omega_m = 0.3$,
$b\rightarrow\infty$) with the one of Born-Infeld electrodynamics
models without cosmological constant ($\Omega_m=1$). Notably the
observations $d_L$ vs $z$ can be well fitted without using
cosmological constant by choosing $\frac{B^2}{2 a_o^{4} b^{2}}\sim
0.05$. In order to better appreciate the role of Born-Infeld
electrodynamics in these curves, Fig.\ref{onlymatter} also
includes the curve resulting from an ordinary (Maxwellian)
cosmology with $\Omega_m=1$.
Although non-linear electrodynamics effects could explain the
curves of luminosity distance vs redshift for type Ia supernovae
\cite{Riess-Perlmutter} without invoking dark energy, one should
be cautious. Accepting a typical value of $10^{-7} G$ \cite{pie3}
for the cosmological background magnetic field $\vert{\bf
B}_o\vert = B/a_o^2$ \cite{galax}, together with the constraint
$b\gtrsim 10^{20} V/m$ for the Born-Infeld parameter \cite{Jack},
then the corrections to the standard cosmology would be
negligible. Even so, non-linear electrodynamics should be
considered as a source of degeneration in the curve $d_L$ vs $z$.
Figure \ref{comparacion} compares the curve of the standard
cosmology ($\Omega_m=0.3$, $\Omega_\Lambda=0.7$, $b\rightarrow
\infty$) with the one resulting from Born-Infeld electrodynamics
for the same values of $\Omega_m$ and $\Omega_\Lambda$. The curves
intersect at $z_r = 0.43$. If $z < z_r$ the curves get their
maximum separation at $z=0.26$. If $z > z_r$ the luminosity
distance predicted by Born-Infeld electrodynamics becomes smaller
than the one from standard cosmology. In this last case the curves
seem to go dramatically apart (however this feature should be
confirmed by extending the calculus up to a higher order of
approximation). The degree of separation of both curves is
governed by the value of $\frac{B^2}{2 a_o^{4} b^{2}}$. Future
observations could allow a test of these features to obtain a
constraint for this value.
\section{Conclusions}
In this paper we have solved the Born-Infeld equations for
electromagnetic plane waves propagating in a background magnetic
field. In the absence of a background field, the Born-Infeld plane
waves are equal to the Maxwell ones. On the contrary, in the
presence of a background magnetic field ${\bf B}$ the non-linear
effects modify both the phase and the amplitude of the wave with
corrections that depend on the combination $\vert{\bf B}\vert^2\,
a^{-4}\, b^{-2}$, where $a$ is the scale factor of the universe.
It is remarkable that Born-Infeld electrodynamics depends on $a$
and $b$ only through the combination $a^4 b^2$. This means that
the Maxwellian approximation ($b\rightarrow\infty$) also
corresponds to the limit $a\rightarrow\infty$. So, although the
electromagnetic field is presently well described by Maxwell
equations for a wide range of phenomena, the non-linear
Born-Infeld electrodynamics could have an influence in the past
when the scale factor was smaller. Therefore the expanding
universe is a good laboratory to test Born-Infeld electrodynamics;
many non-linear aspects of its equations could be relevant when
highly redshifted objects are observed.

In this work we have begun the search for this kind of effects. We
found that the influence of Born-Infeld electrodynamics on the
luminosity distance (\ref{Hod5}) exhibits interesting features
that could be experimentally established by means of more precise
supernova observations and a better knowledge of the cosmological
background fields. Firstly, the experimental data for $d_L$ vs $z$
could be fitted without invoking dark energy, although there is no
observational evidence of the background field that would be
required. Secondly, the shape of the curve $d_L$ vs $z$ predicted
by the standard cosmology ($\Omega_m=0.3$, $\Omega_\Lambda=0.7$,
$b\rightarrow \infty$) for high redshifts differs appreciably from
the one predicted by Born-Infeld electrodynamics, which opens the
possibility of detecting non-linear electrodynamics effects in a
future.

\acknowledgments M.A. and G.R.B. are supported by ANPCyT and
CONICET graduate scholarships respectively. This work was
partially supported by Universidad de Buenos Aires (UBACYT X103)
and CONICET (PIP 6332).


\begin{thebibliography}{}
\bibitem{pie1} In curved space-time,
$T_0^\alpha$ (one covariant index and one contravariant index)
coincide with the tensor components in an orthonormalized basis;
thus, they effectively give the measure of the energy flux.
\bibitem{pie2} The approximation is valid when
$(1+z)^4\,\frac{B_\bot^2}{a_o^4 b^2} \ll 1$; see for instance
(\ref{ao/ae}).
\bibitem{pie3} Notice that the energy density associated to such a field
is much smaller than the matter energy density.
\end{thebibliography}
\end{document}